%% file: template.tex
\RequirePackage{fix-cm}
\documentclass[smallextended]{svjour3}       
\smartqed  
\usepackage{graphicx}
\usepackage{hyperref}
\usepackage{diagbox}
\usepackage{multirow}
\usepackage{booktabs}
\usepackage{tcolorbox}
\usepackage{breakurl}
\usepackage{xurl}
\usepackage[numbers]{natbib}

\begin{document}

\title{Adversarial Robustness of Open-source Text Classification Models and Fine-Tuning Chains}
\subtitle{ - An Empirical Study of Text Classification Models on Hugging Face Hub}

\author{Hao Qin\textsuperscript{*} \and Mingyang Li\textsuperscript{*} \and Junjie Wang \and Qing Wang}

\institute{Hao Qin\textsuperscript{*}, Mingyang Li\textsuperscript{*}, Junjie Wang, Qing Wang are with State Key Laboratory of Intelligent Game, Institute of Software Chinese Academy of Sciences, and University of Chinese Academy of Sciences, Beijing, China.\\
\email{qinhao22@mails.ucas.ac.cn; mingyang2017@iscas.ac.cn; junjie@iscas.ac.cn; wq@itechs.iscas.ac.cn} 
\\\textsuperscript{*}Both authors contributed equally to this work.
}

\date{Received: date / Accepted: date}

\maketitle
\input{0.abstract}

\input{1.introduction}
\input{2.background}
\input{3.sc_construction}
\input{4.rq1}
\input{5.rq2}
\input{6.rq3}
\input{7.implication}
\input{8.threat}
\input{9.rw}
\input{10.conclusion}

\section*{Data Availability Statements}
\sloppy
The models used in this experiment and their upstream model information are available in a GitHub repository. The repository can be accessed through the following link: \url{https://github.com/Xcasdfas/Adversarial-Robustness-of-Open-source-AI-Models-and-Fine-Tuning-Chains}

\section*{Conflict of interest}

The authors declare that they have no conflict of interest.

\bibliographystyle{plainnat}
\bibliography{sample-base}


\end{document}

%% file: 0.abstract.tex
\begin{abstract}

\textbf{Context:}
With the advancement of artificial intelligence (AI) technology and applications, numerous AI models have been developed, leading to the emergence of open-source model hosting platforms like Hugging Face (HF). Thanks to these platforms, individuals can directly download and use models, as well as fine-tune them to construct more domain-specific models. 
However, just like traditional software supply chains face security risks, AI models and fine-tuning chains also encounter new security risks, such as adversarial attacks. 
Therefore, the adversarial robustness of these models has garnered attention, potentially influencing people's choices regarding open-source models.

\textbf{Objective:}
This paper aims to explore the adversarial robustness of open-source AI models and their chains formed by the upstream-downstream relationships via fine-tuning to provide insights into the potential adversarial risks.

\textbf{Method:}
We collect text classification models on HF and construct the fine-tuning chains.
Then, we conduct an empirical analysis of model reuse and associated robustness risks under existing adversarial attacks from two aspects, i.e., models and their fine-tuning chains.

\textbf{Results:}
Despite the models' widespread downloading and reuse, they are generally susceptible to adversarial attack risks, with an average of 52.70\% attack success rate. 
Moreover, fine-tuning typically exacerbates this risk, resulting in an average 12.60\% increase in attack success rates. 
We also delve into the influence of factors such as attack techniques, datasets, and model architectures on the success rate, as well as the transitivity along the model chains.  

\textbf{Conclusions:} 
The results indicate the poor robustness of text classification models on the popular model hosting platform like HF, and raise the awareness of both researchers and model users pay more attention to the security risks of open-source AI models. 
The analysis also provide valuable insights about the adversarial attack manifestation in both single models and fine-tuning chains, potentially facilitate the risk mitigation and defense techniques.  

\end{abstract}

\keywords{Adversarial Attack, Adversarial Robustness, Text Classification Model, Fine-tuning Chain, Hugging Face}

%% file: 1.introduction.tex
\section{Introduction}

In recent years, the rapid development of deep learning technology has significantly propelled the advancement of artificial intelligence (AI), seeing widespread application and garnering immense attention in various fields such as computer vision and natural language processing \cite{Minar2018Recent,Xin2020A}. 
Concomitant with the AI technique, numerous AI models have been widely applied and provide more intelligent and efficient solutions for real-life scenarios \cite{Shinde2023Advancements,Sah2023Analysis}.
However, it is the consensus that building AI models is expensive as the training process typically requires a large number of training data and consumes numerous computing resources, which has been the key challenge in hindering their broad applications \cite{Raju2021Accelerating,bai2022dnnabacus}.

Given the exorbitant costs, increasing users are building the target models by reusing open-source AI models \cite{qi2023reusing,Yang2017Deep}.
With the appropriate open-source models, users can apply them directly or fine-tune them for downstream applications.
Significantly, the fine-tuning technique has gained increasing attention to achieve efficient and effective model building.
Instead of training models from scratch, the fine-tuning technique starts with powerful base models, 
e.g., BERT, GPT, and MAR, pre-trained on extensive corpora \cite{Fu2023A}.
In recent years, fine-tuning from an open-source base model has become the most mainstream paradigm for building specialized models and has seen widespread application across multiple fields \cite{Lin2023Efficient}.
Under the paradigm, an implicit associative relationship forms accordingly, with each base model as an upstream node and the fine-tuned model as the downstream node.

\begin{figure*}[ht]
\centering
\includegraphics[width=0.9\textwidth]{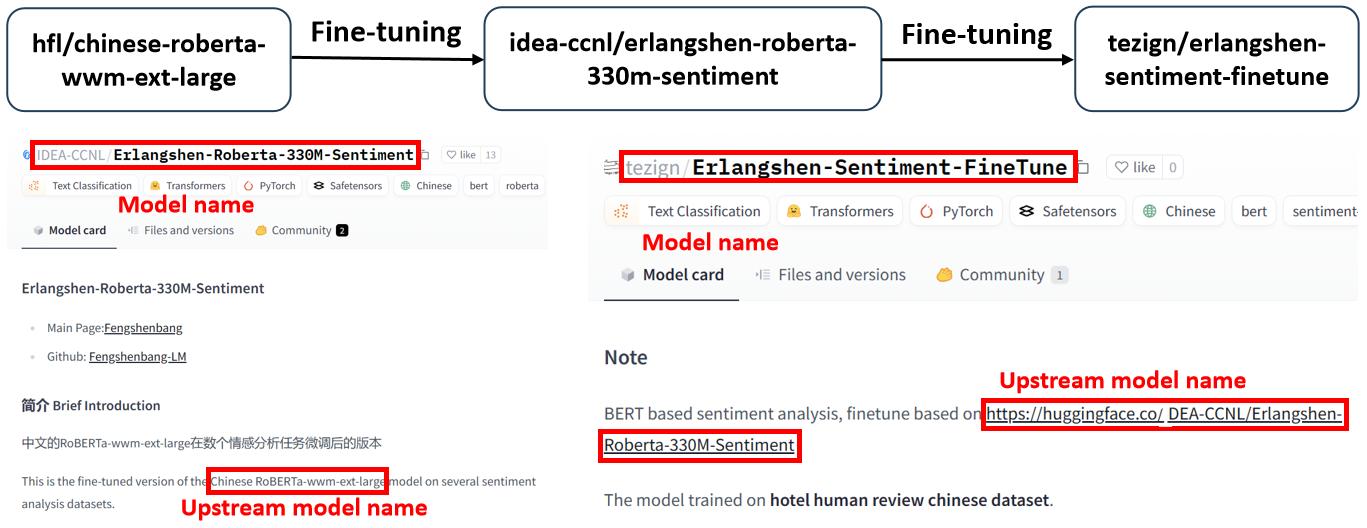}
\caption{Illustrative examples of fine-tuning chain}
\label{fig:fine-tuning chain}
\end{figure*}

The widespread adoption of fine-tuning paradigm and the increasing demand for AI models have led to the emergence of open-source model hosting platforms like Hugging Face (HF).
It is a rapidly growing hub where stakeholders can release, iterate, and share their open-source AI models for model reuse. 
By April 2024, over 5,000 organizations and individuals had contributed 631,866 models and 139,620 associated datasets, including Meta, Google, Microsoft, and Amazon \cite{castaño2024analyzing}.
The open-source models in the HF hubs cover more than forty tasks, including text classification, text generation, and so on.
On HF, besides the general pre-trained models (e.g., BERT \cite{devlin2019bert} and ResNet\cite{he2015deep}), there also exists a multitude of diverse models that are fine-tuned from other AI models.  
Under the fine-tuning paradigm, the upstream and downstream relationships intertwine, thereby forming abundant implicit model chains (as illustrated in Figure \ref{fig:fine-tuning chain}, we name them \textit{fine-tuning chains}).

However, just like the traditional software supply chains face security risks \cite{siadati2024devphish}, the AI models and fine-tuning chains also encounter new security risks, such as adversarial attacks. 
Adversarial robustness, an indicator of the resistance ability to human interference, is a critical factor for ensuring the credibility of AI models in application scenarios \cite{Chen2022Holistic,Madry2017Towards}.
To assess the adversarial robustness, stakeholders typically employ various techniques to generate adversarial samples for attacking,
observing that the model could still make the correct predictions \cite{Chen2019Improving}.
Low adversarial robustness implies that the models are more susceptible to external disturbances, potentially leading to severe erroneous application decisions.
Therefore, it is essential to know the adversarial robustness of the open-source models so that users can be aware of the relevant risks when reusing them. 

Unfortunately, existing studies touched upon aspects such as the reusability of pre-trained models on HF \cite{taraghi2024deep,castaño2024analyzing}, the carbon emissions during model training on HF \cite{Casta_o_2023}, but none of them concern about the security aspects of open-source AI models, particularly in terms of adversarial robustness.
More than that, for the inherent fine-tuning chains buried in model hubs like HF, stakeholders know little about their reuse popularity and the adversarial risks in different reuse scenarios.

To fill this gap, we choose text classification models on HF as subjects because of their popularity and well-studied in terms of adversarial attacks.
We collect 45,688 text classification models from HF and automatically construct their fine-tuning chains 
with the information extraction technique.
After that, we conduct an empirical analysis of the current situation of model reuse and associated adversarial risks from two aspects, i.e., open-source models themselves and their implicit fine-tuning chains.
Accordingly, we answer three research questions as follows.

\textbf{RQ1: General status of open-source text classification models and fine-tuning chains.}
\begin{itemize}
\item RQ1.1: How popular are the open-source text classification models on HF?
\item RQ1.2: How prevalent is the model reuse within HF?
\end{itemize}

RQ1.1 aims to explore the popularity of open-source models through examining their download volumes, while RQ1.2 explores their reuse situation formed by the upstream-downstream relationships via fine-tuning.

\textbf{RQ2. Adversarial robustness of open-source text classification models.}
\begin{itemize}
\item RQ2.1: How robust are the widely-used text classification models under existing adversarial attacks?
\item RQ2.2: Are there any obvious robustness differences for these text classification models under different attack techniques and datasets?
\end{itemize}

This RQ attempts to reveal the model's adversarial robustness under existing adversarial attacks, providing empirical knowledge about the current situation of dominating open-source models for security-sensitive users.
Furthermore, as two key elements of adversarial attacks, we evaluate the adversarial robustness of open-source models  from two perspectives, i.e., attack techniques and datasets.

\textbf{RQ3: Adversarial robustness of open-source fine-tuning chains.}
\begin{itemize}
\item RQ3.1: Are text classification models more robust or more vulnerable to the adversarial attacks after fine-tuning?
\item RQ3.2: Can the vulnerability transfer along the model chains on HF?
\end{itemize}

This RQ aims to investigate the impact of the fine-tuning process on the adversarial robustness of text classification models on HF, respectively from a general perspective (RQ3.1) and more fine-grained perspective of attack samples (RQ3.2), guiding users on the potential threats when reusing open-source models via fine-tuning.

Although 45,688 open source text classification models provide a rich research basis for this study, it is not feasible to conduct a comprehensive experimental analysis due to resource and time constraints. Therefore, this study established a set of strict selection criteria and selected 20 representative models and 10 model chains from a large number of models for in-depth empirical research.
We experiment with six widely-used and state-of-the-art adversarial attack techniques, and four widely-used dataset for conducting attack.
The results show that the open-source text classification models are quite popular in terms of download volume, and prevalent in model reuse within HF.
Despite of that, they are generally susceptible to adversarial attack risks, within an average of 52.70\% attack success rate. 
Furthermore, under different model architectures, the adversarial robustness exhibits different trend after fine-tuning: with the dominating BERT architecture, fine-tuning tends to degrade the robustness of downstream models, resulting in an average of 12.60\% increase in attack success rate; conversely, Electra architecture exhibits an inverse trend.
In addition, the vulnerabilities exhibit a certain degree of transitivity during the fine-tuning process, i.e., 71.3\% samples that successfully attack the upstream model remain effective against the downstream model.

We also explore the implications of our experimental findings, which include providing guidance on selecting suitable methods for reusing models in security-sensitive domains, uncovering potential for more targeted attacks using vulnerable samples from upstream models, drawing inspiration from the Electra architecture to bolster robustness, and more.
This paper highlights the inadequate robustness of text classification models hosted on popular platforms like HF, emphasizing the need for both researchers and model users to pay more attention to the security vulnerabilities in open-source AI models.
Additionally, our analysis offers valuable insights into the occurrence of adversarial attacks across both individual models and fine-tuning chains, potentially aiding in the development of strategies to mitigate risks and implement defense mechanisms.

The key contributions of this study are listed as follows.
\begin{itemize}
    \item 
    To our knowledge, we are the first to investigate the adversarial robustness of the open-source models on HF, particularly the adversarial robustness of the fine-tuning chains.
    \item 
    Our study reveals the inherent robustness risks of open-source text classification models on HF and their transmissibility in the fine-tuning chains. 
    It can provide insights for security-sensitive users to understand the potential risks under different model reuse scenarios.
    \item
    We make the collected open-source models and constructed fine-tuning chains publicly available\footnote{https://github.com/Xcasdfas/Adversarial-Robustness-of-Open-source-AI-Models-and-Fine-Tuning-Chains}, which can be used for replication or reused in future studies.
\end{itemize}

The structure of this paper is as follows: Section 2 introduces the background, Section 3 presents how we collect the models and construct fine-tuning chains. Section 4 to 6 present the experimental design and result analyses for three research questions respectively. Section 7 offers implications of this study, and Section 8 uncovers the threats to validity. Section 9 explores related works, while Section 10 concludes this paper.

%% file: 2.background.tex
\section{Background}

\subsection{Hugging Face (HF)}

Hugging Face (HF) is a company that specializes in natural language processing, an significant branch of AI,  and has become widely recognized for its contributions to the field through its open-source projects and community-driven approach. 
The company is initially and best known for its development of the ``Transformers'' library, which provides a vast array of pre-trained models foundational for various natural language processing tasks, such as text classification, question answering, and machine translation.
This library has garnered a large community of users and contributors, its emergence has simplified access to and deployment of advanced NLP and ML models. Moreover, through its open collaborative community-driven model, it has facilitated the sharing of knowledge and technology on a global scale, making it one of the most influential platforms in advancing natural language processing technology.

Model Hub is the core module for HF that provides a collaborative environment through its online platform.
Researchers, developers, and organizations can upload and share their pre-trained models and datasets, making them accessible to the broader community.
The Hub supports various machine learning frameworks and provides tools for version control, model hosting, and easy integration into existing projects. 
This collaborative environment fosters innovation and accelerates the development of new AI applications.
By April 2024, over 5,000 organizations and individuals had contributed 631,866 models and 139,620 associated datasets, covering more than forty tasks including text classification, text generation, and so on.

\subsection{Adversarial Attacks And Adversarial Robustness Assessment}
\label{sec:adversarial_attack}

Adversarial attacks against AI models represent a critical and burgeoning area of study within the field of AI security. 
These attacks involve the crafting of input data that is slightly altered from the original samples but done so in a way that is imperceptible or minimally perceptible to humans. Despite these minor alterations, such adversarially modified inputs can cause an AI model to make incorrect predictions or classifications with high confidence. 
The susceptibility of AI models to these insidious attacks highlights significant vulnerabilities in their ability to generalize from the training data to real-world scenarios. 
This phenomenon raises profound security concerns, particularly in applications where the integrity and reliability of model predictions are paramount, 
For example, in the field of autonomous driving~\cite{kim2019grounding}, adversarial attacks can cause the vehicle perception system to misrecognize traffic signs, leading to incorrect driving decisions and increasing the risk of traffic accidents. In the context of fake news detection~\cite{Shreyash2022FACTIFY}, adversarial attacks can deceive detection systems, preventing them from timely identifying and filtering false information, thereby accelerating the spread of misinformation, misleading public opinion, and potentially destabilizing society. In medical auxiliary diagnosiss~\cite{YanakaNCK23}, adversarial attacks can cause AI diagnostic systems to misclassify or predict patient data, resulting in incorrect diagnoses and treatment plans, severely endangering patient health and safety.
Understanding and mitigating the effects of adversarial attacks is thus of paramount importance, necessitating a deeper exploration of model robustness and the development of techniques to safeguard against such vulnerabilities.

\begin{figure}[ht]
\centering
\includegraphics[width=\columnwidth]{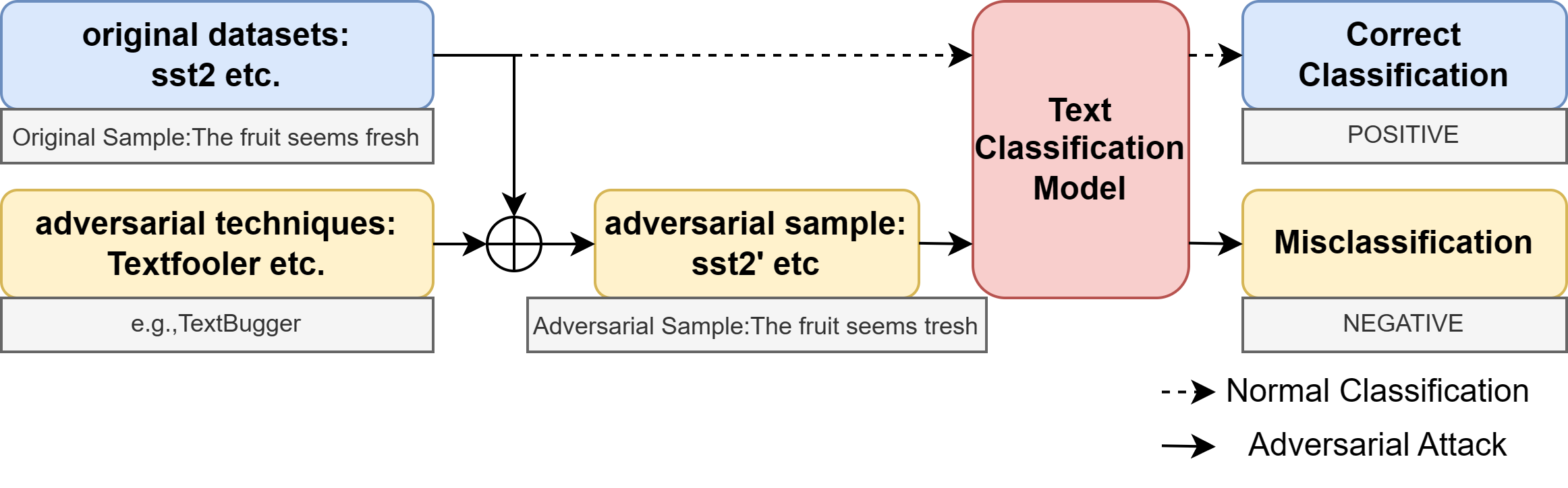}
\caption{The general framework for adversarial attack techniques}
\label{fig:Adversarial attack flow chart}
\end{figure}

To assess the adversarial robustness of AI models, developers typically employ some existing adversarial attack techniques, such as TextBugger~\cite{li2018textbugger} and HotFlip~\cite{ebrahimi2018hotflip} for attacking.
Figure \ref{fig:Adversarial attack flow chart} shows the general framework for adversarial attack techniques.
For NLP tasks, adversarial attacks generally start with an \textit{original dataset}.
For each sample in the original dataset (namely \textit{Original Sample}), attack techniques identify vulnerable characters, words, or entities within the input text based on the feedback (e.g., gradient, logits, or probabilities) from the AI model under assessment.
After that, the attack techniques generate new samples, known as \textit{adversarial samples}, through perturbing the vulnerable elements via character insertion, word substitution, etc.
Then, the generated adversarial samples are sent to the AI models under assessment.
The attack is considered successful if the assessed model produces the incorrect output.
sarial attacks is shown in Figure \ref{fig:Adversarial attack flow chart}.

%% file: 3.sc_construction.tex
\section{Model Collection and Fine-tuning Chain Construction}

\subsection{Model Collection}
\label{sec:data_collection}
To address the research questions, we collect the open-source models for text classification on HF.
Following the data collection in the previous study~\cite{castano2023analyzing}, we employ the HfApi library\footnote{https://github.com/huggingface/hfapi/}, a Python wrapper of HF Hub API, to collect the stored open-source models and their associated information.
Based on the upload timestamp of the stored models, we collected all text classification models before January 12, 2024.
For each model, we obtained information across five dimensions, i.e., \textit{Model Name}, \textit{Model Card}, \textit{Upstream}, \textit{Dataset}, and \textit{Downloads}.
\textit{Model Name} is the unique identifier for each model on HF through which users can directly access the inference interface and obtain the prediction result for the input.
\textit{Model Card} is the basic description in unstructured text, typically documenting how the model is built, how to use it, performance metrics such as classification accuracy, and other fundamental information.
\textit{Upstream} and \textit{Dataset} are optional fields representing the upstream model and the dataset used during the training or fine-tuning process when building the current model.
\textit{Downloads} count the number of times the model has been downloaded for the past month on HF.
Finally, we obtain 45,688 open-source models for text classification on HF, and the detailed attributes for each model are given in our public repository.

\subsection{Fine-tuning Chain Construction}
\label{sec:chain_cunstruction}

Given the collected models, we further construct the fine-tuning chains.
We first identify the upstream model for each collected model.
Ideally, the upstream for each model is recorded through the ``Upstream'' attribute, with which the upstream-downstream relationship can be easily identified by parsing this field.
However, since this attribute is not mandatory for the developers, it is mostly empty on HF, even though an upstream model exists.
Generally, developers mention corresponding upstream models through the model descriptions in the ``Model Card'' attribute.

\begin{figure}[htbp]
\centering
\includegraphics[width=\columnwidth]{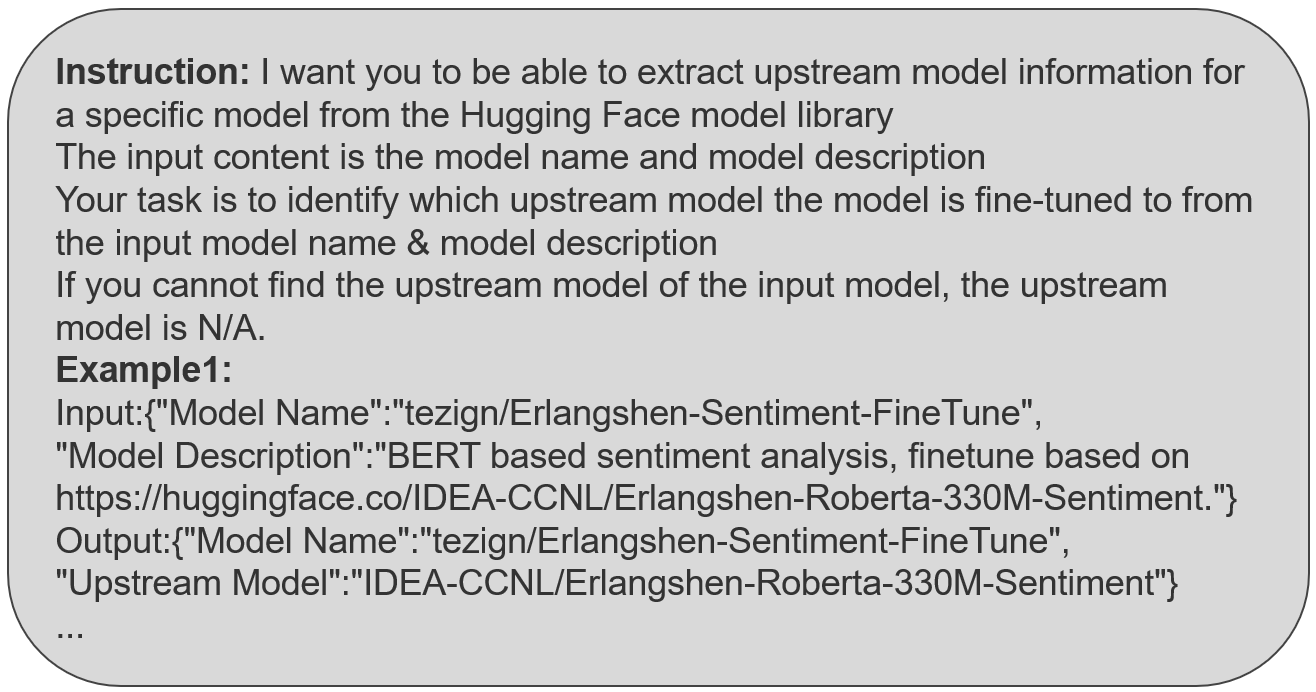}
\caption{The prompt for identifying the upstream model name from the descriptions in ``Model Card''}
\label{fig:prompt}
\end{figure}

Based on this situation, we integrate the ``Upstream'' and ``Model Card'' attributes to identify upstream models.
Specifically, we first check a model's ``Upstream'' attribute. 
If it is not empty, we use this attribute as the index to retrieve the upstream model by matching each collected model's ``Model Name''.
Otherwise, we utilize the descriptions in the ``Model Card'' to identify the name of the upstream model for matching.
Considering that the names of upstream models are typically entities within the complex unstructured texts (as shown in Figure \ref{fig:fine-tuning chain}), and traditional regular expression methods are ineffective for extracting such information or involving a large amount of labeled data for model training, we utilize ChatGPT\footnote{https://openai.com/blog/chatgpt}, a popularly-used large language model (LLM),
guiding it with a carefully crafted prompt to extract the names of upstream models.
For a collected model, if there is no upstream model name extracted from the model descriptions, it is conisidered an isolated node recoreded in our dataset.
Figure \ref{fig:prompt} shows the crafted prompt, where the ``Instruction'' gives the task description and ``Example'' guides the LLM to understand the task it is dealing with and the corresponding input-output format through specific examples.

Regarding the bias introduced by ChatGPT, we manually evaluate the performance of identifying upstream model names. 
Initially, we randomly selected 100 models that are not annotated with upstream models in the ``Upstream'' attributes.
Three researchers manually annotate the upstream model names for each model respectively and establish a ground truth through discussion. 
Based on this, we evaluated the proportion of correct identification, i.e. accuracy rate, and the results showed an accuracy rate of 97\% on the 100 samples, demonstrating promising reliability of our automatic upstream model identification.

After that, we match the identified results with the ``Model Name'' of other collected models to identify the upstream-downstream relationships.
By integrating all identified upstream-downstream relationships, we obtained the fine-tuning chains.
Finally, we built 29,148 fine-tuning chains, involving 31,672 different models. 
Based on the collected models and constructed fine-tuning chains, we conduct the empirical study to answer the three research questions.

%% file: 4.rq1.tex
\section{Answering RQ1}

\subsection{Experiment Design}

This RQ mainly focuses on two aspects: the popularity of text classification models on HF, indicated by the number of downloads (RQ1.1), and their reuse situation within HF, indicated by the frequency serving as the upstream models (RQ1.2).
For RQ1.1, given collected models (details in Section \ref{sec:data_collection}), we parse the ``Downloads'' attribute and rank the models according to the number of downloads in descending order, and we use it as an indicator for the model popularity.
We choose the number of downloads as indicator as downloading behavior is typically an initial indication of user interest and potential further use of the model.
Additionally, we focus on the models with the most downloads and analyze the proportion of the top K downloaded models to the total downloads of all collected models.
These statistics will help reveal the distribution of model downloads across HF and imply their popularity.  
For RQ 1.2, We analyze the proportion of models that mention upstream models in the corresponding ``Upstream''  or ``Model Card'' attribute, to assess the prevalent of model reuse on the HF.
Regarding the constructed model chains, we further analyze their characteristics, e.g., chain length, the most frequent upstream models, reuse counts, etc.

\subsection{Results And Discussions}

\subsubsection{\textbf{RQ1.1: How popular are the open-source text classification models on HF?}}
\label{sec:rq1.1_result}
\begin{figure}[htbp]
\centering
\includegraphics[width=\columnwidth]{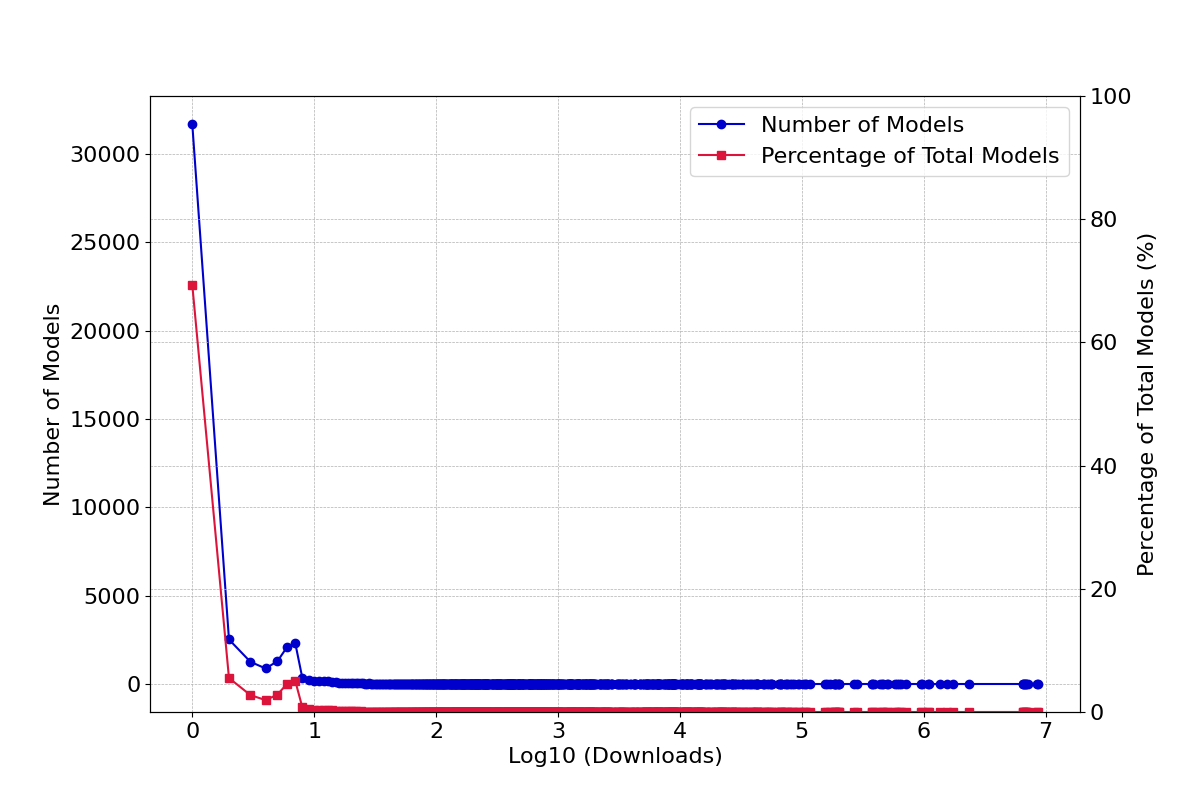}
\caption{The download distribution of the text classification models on HF
(RQ1)}

\label{fig:model_downloads}
\end{figure}

Figure \ref{fig:model_downloads} shows the download distribution of the text classification models of HF.
In general, numerous text classification models on HF have broadly received much attention, and 38.63\%
of the 45,588 models are downloaded at least once within a monthly cycle.
In addition, 1,536 model (about 3.36\% of all the collected models) have gained significant interests with more than 30 downloads within a monthly cycle. 
The results highlight the popularity of the various text classification models on HF.

Furthermore, we also observe the long-tail effect in the downloads on HF.
Specifically, the top 20 models alone account for 72,967,027 downloads, representing 86.10\% of the total number. 
About top 0.1\% of the models (46 models) account for 94.85\% of the total downloads.
These results also reveal that, on the fact that models receive widespread attention on HF, users exhibit a certain degree of preference for a small number of top models.
For the models receiving much attention, it is non-trivial to conduct rigorous quality assessments to ensure their reliability.

\begin{tcolorbox}[colback=gray!50!white]
\textbf{Finding 1}:
\textit{
Over 38.63\% text classification models are downloaded at least once within a monthly cycle, highlighting the widespread attention of the models on HF.
Furthermore, the distribution of downloads on HF demonstrates a long-tail effect that a few models at the top account for considerable downloads.
This situation reveals their gained preference and the necessity for a more stringent assessment of their quality features (e.g., adversarial robustness).
}
\end{tcolorbox}

\subsubsection{\textbf{RQ1.2: How prevalent is the model reuse within HF?}}

By identifying the upstream for each model (details in Section \ref{sec:chain_cunstruction}), we found that about 64.66\% of the collected models have the corresponding upstream models.
This indicates that models of a certain scale are built by reusing other open-source models, and substantial models along with their upstream are involved in the fine-tuning chains on HF.
In addition, out of the 45,688 text classification models, 455 models are used as upstream models, accounting for 1\% of the total.
We found that most of them are large-scale trained models released by companies or organizations such as CardiffNLP and Microsoft.
The results imply that developers are inclined to reuse such famous models rather than the unfamous ones. 

\begin{figure}[htbp]
\centering
\includegraphics[width=\columnwidth]{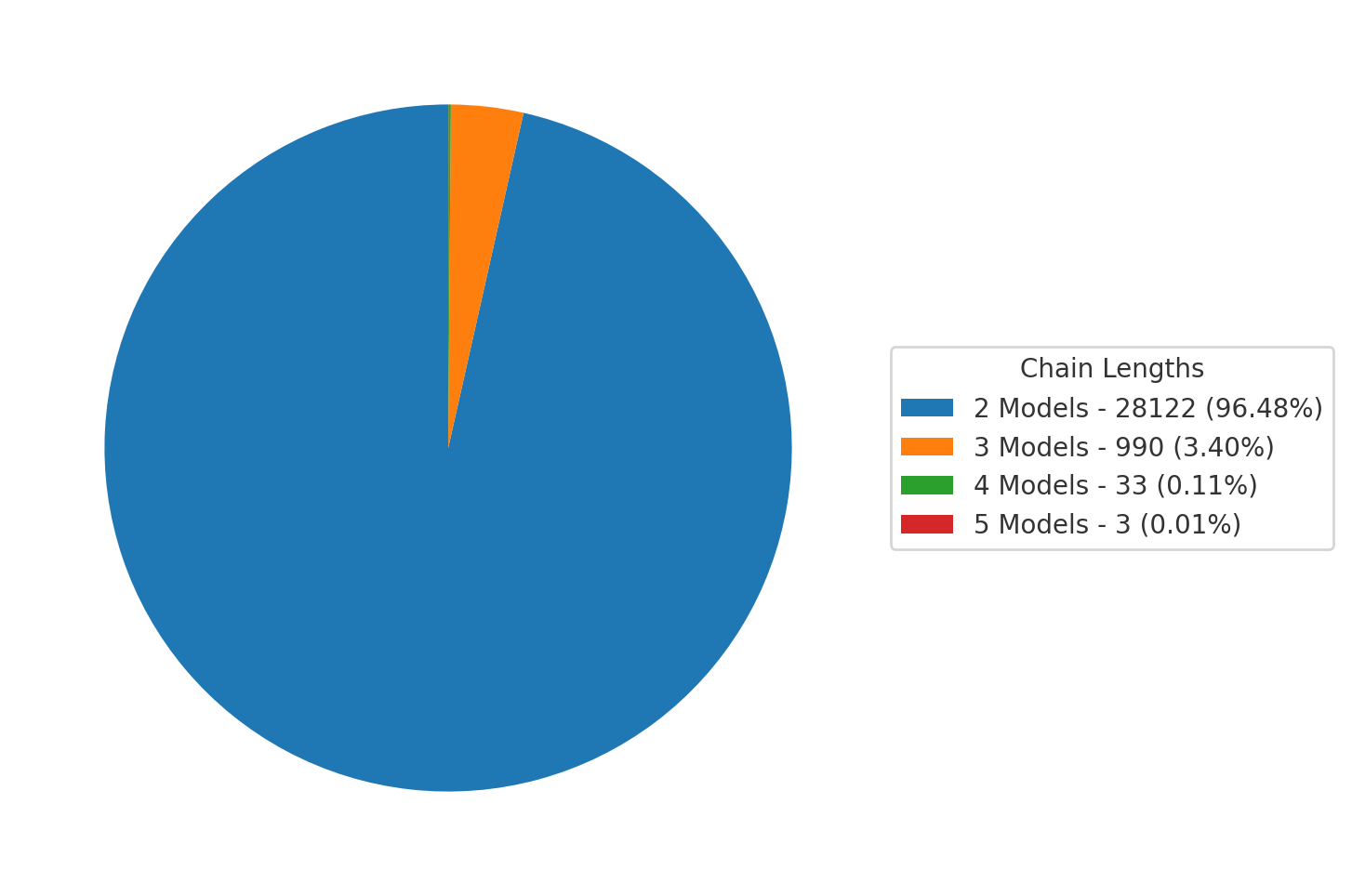}
\caption{Distribution of model chain lengths (RQ1)
}
\label{fig:Distribution of the number of models on the model chain}
\end{figure}

For the model lengths, Figure \ref{fig:Distribution of the number of models on the model chain} shows the distribution of the number of models on each model chain.
Overall, among the 29,153 model chains collected, there are an average of 2.04 models on each chain, which is significantly shorter than the length of traditional open-source software supply chains \cite{wang2024large}.
Specifically, chains primarily consist of two or three models (proportions sum up to 99.88\%). 
The proportion of model chains made up of two models accounts for 96.48\%, and those consisting of three models account for 3.40\%. 
The shorter length of these model chains may be due to the fact that most text classification models are fine-tuned from large-scale  models such as BERT and DistilBERT. 
Given the high performance of these base models, developers usually only need one or two rounds of fine-tuning to achieve the desired results, eliminating the necessity for multiple fine-tuning sessions.

Given the above circumstance, we further analyzed which models are more frequently used as upstream models for text classification.
Generally, model reuse is prevalent and 1\% model are reused at least once with HF.
Table \ref{tab:upstream_model} shows the top 10 most popular models used as the upstream ones in the model chains on HF, 
Where the "Downstream (\%)" column indicates the number of downstream models fine-tuned by the current model and the corresponding proportion of all upstream-downstream model pairs where the model is identified as upstream, 
``Downstream Task'' indicates the downstream task where upstream models are most commonly applied, and ``Downloads (Ranking)'' shows the numbers of downloads and rankings of these models.
According to ``Downstream (\%)'', the top 10 (2.20\% of all reused text classification models) most popular upstream models contribute 30.93\% of model reuse for text classification on HF. 
The result is similar to the analysis in download volume (details in Section \ref{sec:rq1.1_result}), both conforming to a certain degree of the long-tail effect.
This phenomenon also emphasizes the importance of assessing the reliability of a few core models, as their potential vulnerabilities could significantly impact a wide range of downstream applications.

\input{RQ1_Upstream_model_in_text_classification_model}

\begin{tcolorbox}[colback=gray!50!white]
\textbf{Finding 2}:
\textit{
Substantial models for text classification are built by reusing the existing open-source models, which forms fine-tuning chains of considerable scale within HF. 
Furthermore, developers of downstream models are more inclined to choose a few famous models for reuse, which emphasizes the quality of these core models.
}
\end{tcolorbox}

%% file: RQ1_Upstream_model_in_text_classification_model.tex
\begin{table*}[htbp]
\small
  \caption{The most popular upstream models in the model chains on HF (RQ1)}
  \label{tab:upstream_model}
  \resizebox{\textwidth}{!}{
  \begin{tabular}{cccc }
    \toprule
    \textbf{Model Name} & \textbf{Downstream Task} &   \textbf{\#Downstream (\%)} & \textbf{\#Downloads (Ranking)}\\
    \midrule
    distilbert-base-uncased-finetuned-sst-2-english & general sentiment analysis & 137 (10.75\%) & 6849685 (5)\\
    cardiffnlp/twitter-roberta-base-sentiment-latest & general sentiment analysis & 45 (3.53\%) & 8622648 (1)\\
    cardiffnlp/twitter-roberta-base-sentiment & general sentiment analysis & 38 (2.98\%) & 1744622 (12)\\
    nlptown/bert-base-multilingual-uncased-sentiment & product review sentiment analysis & 33 (2.59\%) & 1552518 (13)\\
    microsoft/minilm-l12-h384-uncased & general sentiment analysis & 32 (2.51\%) & 12363 (133)\\
    prosusai/finbert &  financial sentiment analysis & 26 (2.04\%) & 1368118 (14)\\
    roberta-large-mnli & textual entailment & 23 (1.81\%) & 95071 (46)\\
    cardiffnlp/twitter-xlm-roberta-base-sentiment & general sentiment analysis & 21 (1.65\%) & 1084083 (16)\\
    papluca/xlm-roberta-base-language-detection & language detection & 20 (1.57\%) & 502579 (27)\\
    siebert/sentiment-roberta-large-english & general sentiment analysis & 19 (1.49\%) & 117489 (44)\\
    \bottomrule
  \end{tabular}
}
\end{table*}

%% file: 5.rq2.tex
\section{Answering RQ2}

\subsection{Experiment Design}

\subsubsection{Chain and Model Selection}
\label{sec:chain_selection}

To investigate the adversarial robustness of open-source models (RQ2) and robustness change during model fine-tuning (RQ3), we conduct chain and model selection.
First, of all the constructed model chains, we filter them by following criteria: 
(1) all the datasets for model training or fine-tuning on the chain should be declared to guarantee the selected subjects are of higher description quality, and (2) all the models on the chain should own at least 30 downloads to ensure that the subjects have a certain level of popularity.
After that, we obtained ten upstream-downstream model pairs, and 18 open-source models were involved (3 pairs shared the same upstream model). 
Table \ref{Model Information} details the selected chains and involved models.
In addition, based on the results for RQ1, we additionally introduce the model with the most downloads (mrm8488/distil\-roberta-finetuned-financial-news-sentiment-analysis) and the model with the most reuse (distilbert-base-uncas\-ed-finetuned-sst-2-english) on HF. 
Despite the fact that the model chains derived from cardiffnlp/twitter-roberta-base-sentiment-latest and mrm8488/distilroberta-finetuned-financial-news-sentiment-analysis do not meet the selection criteria of this study, we decided to include these models in our research due to their top positions in terms of downloads and reuse on the Hugging Face platform. This decision was made to evaluate the security of these models, given their high practical application value.
Finally, we obtained 20 models and ten upstream-downstream model chains to investigate their adversarial robustness for RQ2 and RQ3.

\subsubsection{Adversarial Robustness Assessment for Models}
We employ six widely-used and state-of-the-art adversarial sample generation techniques for attacking to investigate the adversarial robustness, as shown below. 

\begin{itemize}
    \item 
    \textbf{TextBugger~\cite{li2018textbugger}.} 
    It first adopts a scoring mechanism to determine the importance of words or characters in the text based on their impact on the model’s output. Then, it employs a series of perturbation techniques to generate adversarial samples for attacking, such as character insertion, deletion, swapping, or word substitution, targeting these critical elements.

    \item 
    \textbf{HotFlip~\cite{ebrahimi2018hotflip}.} 
    It employs the model’s gradients to determine which characters or words, when altered, will have the most significant impact on the model’s decision. 
    HotFlip then applies these perturbations to generate adversarial samples - which can include flipping characters and inserting or deleting them - to minimize changes to the original input while maximizing the likelihood of fooling the model into making incorrect predictions.

    \item 
    \textbf{TextFooler~\cite{jin2020bert}.} 
    First, it identifies the most critical words in the input text by assessing changes in the output confidence as each word is removed or altered. 
    Next, TextFooler searches for semantically similar but syntactically different substitutes for these critical words to find replacements that maintain the original meaning as closely as possible. 
    Lastly, it evaluates the new text to ensure that the substitutions not only misled the targeted model into a wrong prediction but also preserved the original text's grammatical correctness and semantic coherence, thereby keeping changes imperceptible to human readers.

    \item 
    \textbf{PWWS~\cite{ren2019generating}.}
     Initially, PWWS calculates the word saliency by modifying or removing each word and observing the change in the model’s output.
     PWWS then seeks to find appropriate replacements for these words based on their semantic similarity, aiming to preserve the overall meaning of the text. 
     Finally, PWWS re-evaluates the adversarial text to ensure that the changes are not only effective at deceiving the model but also subtle enough to appear natural and coherent to human readers.
    
    \item 
    \textbf{SCPN~\cite{iyyer2018adversarial}.} 
    Firstly, SCPN identifies target sentences or phrases within the text that are important for the model’s prediction. 
    It then uses its trained paraphrase model to generate alternatives to these sentences that are semantically equivalent but lexically different. 
    
    \item
    \textbf{GAN~\cite{zhao2017generating}.}
    GAN hinges on a duel between two neural networks: a generator and a discriminator. The generator creates data instances that mimic the true data distribution, aiming to fool the discriminator, which is trained to distinguish between the generator’s fake instances and real data. 
    Through this adversarial training process, the generator are guided to generate adversarial samples that are close to the original yet modified subtly to cause misclassification by the target model.
\end{itemize}

These attack techniques are designed according to  various technical principles and could be used to evaluate the adversarial robustness of AI models from different aspects. 
Specifically, TextBugger and HotFlip represent character-level attacks, generating adversarial samples through character insertion, deletion, and substitution. TextFooler and PWWS represent word-level attacks, achieving adversarial effects through word replacement. SCPN represents sentence-level attacks, generating semantically equivalent but syntactically different sentences to confuse the model. GAN represents generative adversarial network attacks, creating samples close to the original but subtly modified to deceive the discriminator. This diverse set of attack methods ensures that the experiments cover various types of adversarial attacks, providing a robust assessment of the models' performance under different adversarial environments.

In addition, we introduced four widely used datasets, i.e., sst2 \cite{socher-etal-2013-recursive}, IMDB \cite{maas-EtAl:2011:ACL-HLT2011}, rotten tomatoes \cite{Pang+Lee:05a}, and Amazon polarity \cite{zhang2016characterlevel} as the original datasets (the terminology is described in Section \ref{sec:adversarial_attack}) for adversarial sample generation. 
These datasets are commonly used in NLP tasks and encompass various text lengths and language styles, ensuring the broad applicability and representativeness of the experimental results. 
In this study, we sampled 100 instances from the training sets of each of these four datasets to generate adversarial samples and evaluate the robustness of the models.
With the original datasets, we use OpenAttack~\cite{Zeng_2021}, which is the open-source implementation of the above techniques, for adversarial attacks.
We obtained the output for each generated adversarial sample through the inference interface provided by HF and observed whether the subject model produced the incorrect prediction.
Then we use the attack success rate (ASR), a commonly used measure \cite{dang2024curious} referring the proportion of all the adversarial samples where models give the incorrect prediction, as the metric to explore the adversarial robustness under different attack techniques and different original datasets.

\input{RQ2_Model_info}

\subsection{Results And Discussions}

\subsubsection{\textbf{RQ2.1: How robust are the widely-used text classification models under existing adversarial attacks?}}

\begin{figure}[htbp]
\centering
\includegraphics[width=\columnwidth]{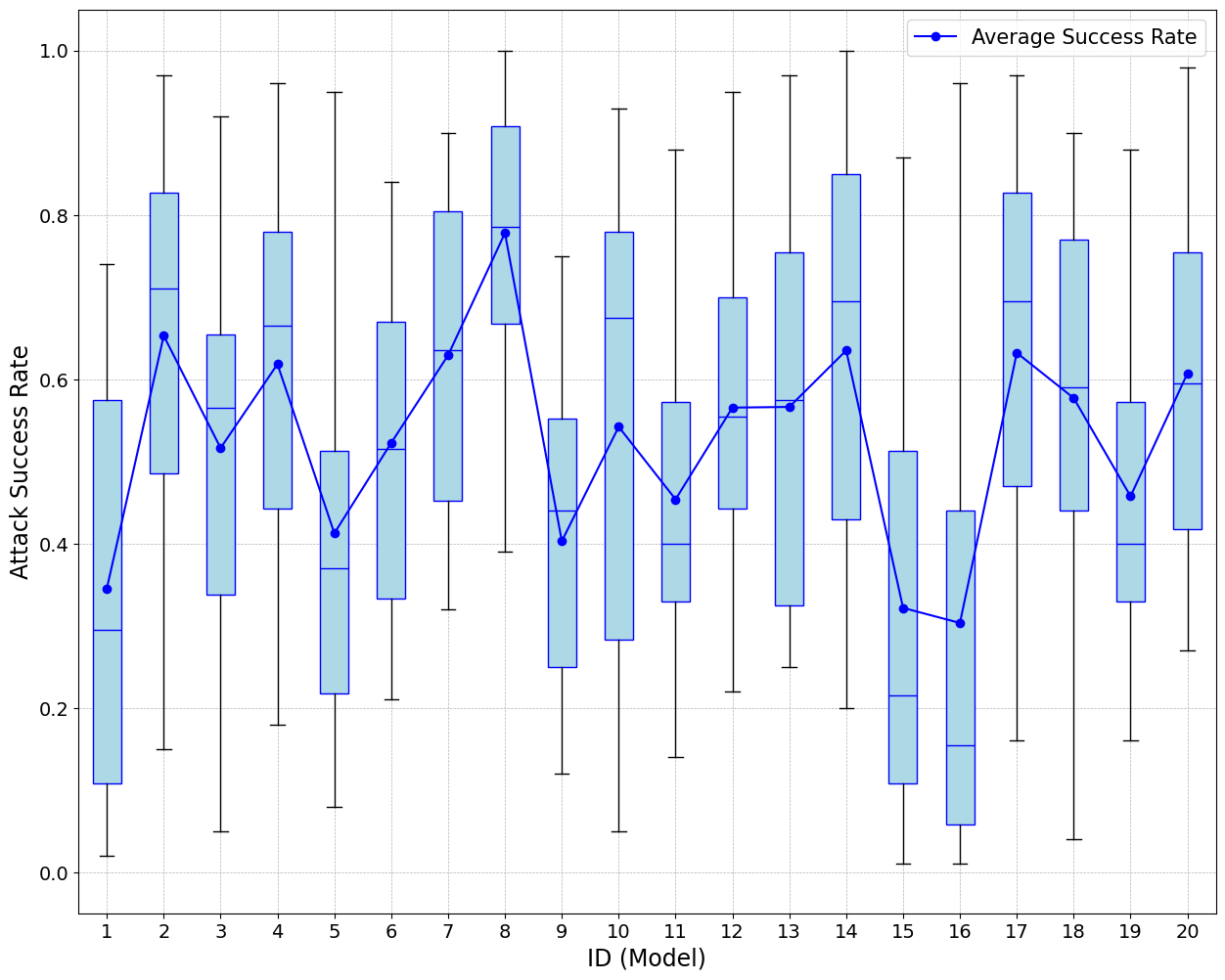}
\caption{The average ASRs for the text classification models (RQ2)}
\label{fig:asr_overall}
\end{figure}

Figure \ref{fig:asr_overall} shows the average ASR of each text classification on HF under different attack approaches using the four datasets.
In general, the ASRs are not low for all models, ranging from a low of 30\% to a high of 78\%.
The results indicate that the adversarial robustness is unsatisfactory for all the studied models and the models suffer adversarial risks when directly reusing them in application scenarios.

\subsubsection{\textbf{RQ2.2: Are there any obvious robustness differences for these text classification models under different attack techniques and datasets?}}

\begin{figure}[ht]
\centering
\includegraphics[width=\columnwidth]{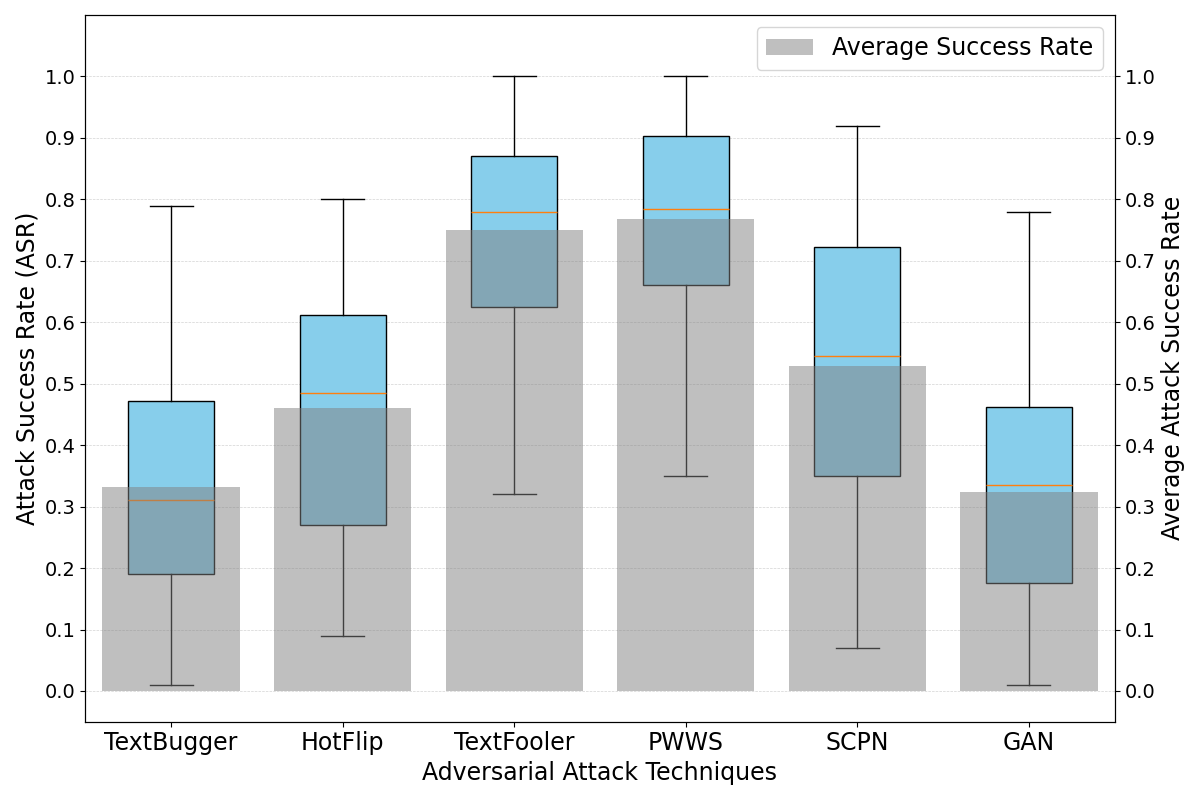}
\caption{The ASRs of widely-used models for text classification under different adversarial attack techniques (RQ2)}
\label{fig:asr_approaches}
\end{figure}

Figure \ref{fig:asr_approaches} shows the average ASRs of different adversarial attack methods on twenty models and four datasets. 
Experimental results show that using different adversarial attack methods significantly impacts the average ASR. 
Among them, the TextFooler and PWWS methods show the highest ASR, which indicates that the model is vulnerable to attacks based on synonym substitution. 
The lower success rates of TextBugger and GAN mean that the model is better resistant to these attacks. 

\begin{figure}[ht]
\centering
\includegraphics[width=\columnwidth]{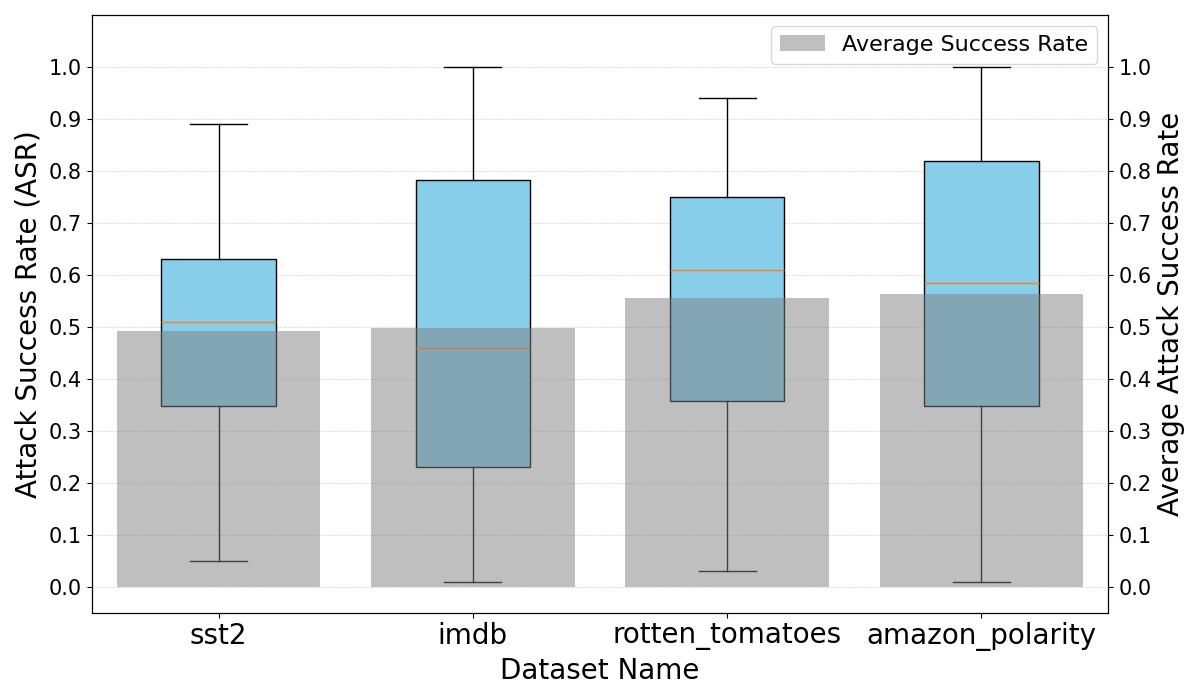}
\caption{The ASRs of widely-used models for text classification under different datasets (RQ2)}
\label{fig:asr_dataset}
\end{figure}

Figure \ref{fig:asr_dataset} shows the average ASRs of different datasets on the wide-used models for text classification. 
The experimental results indicate that the choice of dataset has limited impact on the average ASR of the adversarial attacks, although there are significant differences in belonging domain, language styles and terminologies among the four datasets.
Furthermore, considering the robustness risks associated with the models, security-sensitive users can consider designing targeted defense methods (such as adversarial sample detection \cite{wang2018detecting,Wang_2019}, adversarial training \cite{shafahi2019adversarial,Kurakin2016Adversarial}, etc.) based on the principles and application scope of the attack techniques, to enhance the security in real-world scenarios.

\begin{tcolorbox}[colback=gray!50!white]
\textbf{Finding 3}:
\textit{
The widely used models for text classification exhibit non-negligible attack vulnerability (30\% - 78\% ASR), indicating the low attack robustness of open-source AI models.
In addition, the attack techniques have a noticeable impact on the ASRs of these models.
Users who want to reuse open-source models directly in their application scenarios may face a certain degree of robustness risk.}
\end{tcolorbox}

%% file: RQ2_Model_info.tex
\begin{table*}[htbp]
  \caption{The subject fine-tuning chains and involved models for adversarial robustness assessment (RQ2)}
  \label{Model Information}
  \resizebox{\textwidth}{!}{
  \begin{tabular}{c cccc }
    \toprule
    Chain ID& Model ID & Model Name & Downloads & Architecture\\
    \midrule
    \multirow{2}{*}{C1} & M1 & mrm8488/distilroberta-finetuned-tweets-hate-speech & 275 & BERT \\
                        & M2 & hackathon-pln-es/detect-acoso-twitter-es & 219 & BERT \\
    \multirow{2}{*}{C2} & M3 & juliensimon/reviews-sentiment-analysis & 1819 & BERT \\
                        & M4 & houssemmammeri/revsen-v1 & 122 & BERT \\
    \multirow{2}{*}{C3} & M5 & ProsusAI/finbert & 1368118 & BERT \\
                        & M6 & ziweichen/finbert-fomc & 203 & BERT \\
    \multirow{2}{*}{C4} & M5 & ProsusAI/finbert & 1368118 & BERT \\
                        & M7 & kk08/cryptobert & 798 & BERT \\
    \multirow{2}{*}{C5} & M5 & ProsusAI/finbert & 1368118 & BERT \\
                        & M8 & yiyanghkust/finbert-esg & 2583 & BERT \\
    \multirow{2}{*}{C6} & M9 & idea-ccnl/erlangshen-roberta-330m\-sentiment & 1819 & BERT \\
                        & M10 & tezign/erlangshen-sentiment-finetune & 64 & BERT \\
    \multirow{2}{*}{C7} & M11 & mrm8488/distilroberta-finetuned-financial-news-sentiment-analysis & 5732795 & BERT \\
                        & M12 & finscience/fs-distilroberta-fine-tuned & 68 & BERT \\
    \multirow{2}{*}{C8} & M13 & jarvisx17/japanese-sentiment-analysis & 1762 & BERT\\
                        & M14 & minutillamolinara/bert-japanese\_finetuned-sentiment-analysis & 186 & BERT\\
    \multirow{2}{*}{C9} & M15 & gooohjy/suicidal-electra & 1338 & Electra\\
                        & M16 & sentinet/suicidality & 613 & Electra\\
    \multirow{2}{*}{C10} & M17 & hazqeel/electra-small-finetuned-malay-english & 114 & Electra\\
                         & M18 & hazqeel/electra-small-doa-finetuned-ms-en-v3 & 115 & Electra\\
    \bottomrule
  \end{tabular}
}
\end{table*}

%% file: 6.rq3.tex
\section{Answering RQ3}

\subsection{Experiment Design}

Given the 10 selected model chains (details in Section \ref{sec:chain_selection}), we further explore the impact of fine-tuning on the adversarial robustness, aiming to guide the developers with the adversarial risks when fine-tuning the open-source models on HF for model reuse.

For RQ 3.1, we calculated the average ASR of adversarial attacks on upstream models across six attack approaches powered with four datasets (denoted as $ASR_{up}$), as well as the average ASR for the corresponding downstream models (denoted as $ASR_{down}$). 
After that, we calculated $\delta=ASR_{down}-ASR_{up}$, which is the difference between the average success rates of attacks on downstream models and their upstream counterparts.
If $\delta$ is positive, it indicates that the downstream model is more susceptible to attacks compared to the upstream model, suggesting that the fine-tuning process may have reduced robustness; otherwise, it indicates an improvement in the robustness of the downstream model.

For RQ 3.2, for an original sample that successfully attacks the upstream model, i.e. vulnerable sample, we explore whether it still successfully attacks the corresponding downstream model.
To quantify this transmissibility, we defined ``transferable rate'' as the proportion of samples that successfully attacked the upstream model and were also successful in attacking the downstream model.
Specifically, we use different adversarial attack techniques equipped with different original datasets to generate adversarial samples. 
For each upstream-downstream pair involved in 10 model chains, we identify vulnerable samples through the upstream model and examine whether they could successfully attack the corresponding downstream model.
In the end, we calculate their average ``transferable rate'' for each chain separately.

\subsection{Results And Discussions}

\subsubsection{\textbf{RQ3.1: Are text classification models more robust or more vulnerable to common adversarial attacks after fine-tuning?}}
\label{subsubsec_RQ3.1}
\begin{figure}[htbp]
\centering
\includegraphics[width=\columnwidth]{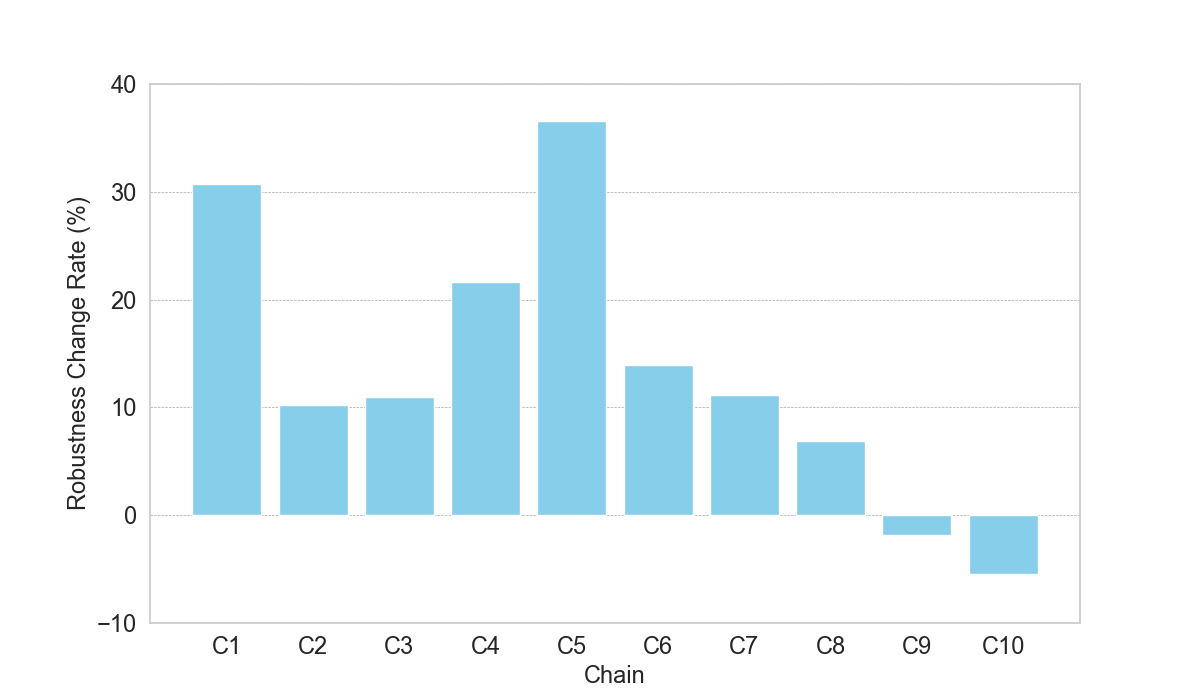}
\caption{The change in adversarial robustness before and after fine-tuning for each upstream-downstream pair (RQ3)}
\label{fig:robustness_change}
\end{figure}

Figure \ref{fig:robustness_change} shows the average value of $\delta$ across 24 different attack scenarios (covering six attack approaches and four original datasets). 
For C1-C8, the $\delta$ is positive, indicating they exhibit consistent changes in adversarial robustness before and after fine-tuning, with the downstream model becoming less robust (average $delta$ is 16.84\%).
Especially for C1 and C5, their adversarial robustness experiences the most significant declines, exceeding 30\% and 35\% in $\delta$, respectively.
However, for the other two (C9, C10), their robustness improves.   
Further analysis revealed that all models in the C1-C8 follow the BERT architecture \cite{devlin2019bert}, while C9 and C10 belong to Electra architecture \cite{clark2020electra}, which will be discussed later.

\begin{tcolorbox}[colback=gray!50!white]
\textbf{Finding 4}:
\textit{
After fine-tuning, whether robustness improves or declines is largely related to the architecture the model employs. 
For the currently dominant BERT architecture, its model robustness tends to decrease compared to the upstream model. 
While, in specific architectures such as Electra, empirical evidence suggests that their robustness does not decrease but enhances, thus rendering it particularly advantageous for security-conscious users.}
\end{tcolorbox}

\subsubsection{\textbf{RQ3.2: Can the vulnerability embodying in vulnerable samples transfer along the model chains on HF?}}
\label{subsubsec_RQ3.2}

\begin{figure}[htbp]
\centering
\includegraphics[width=\columnwidth]{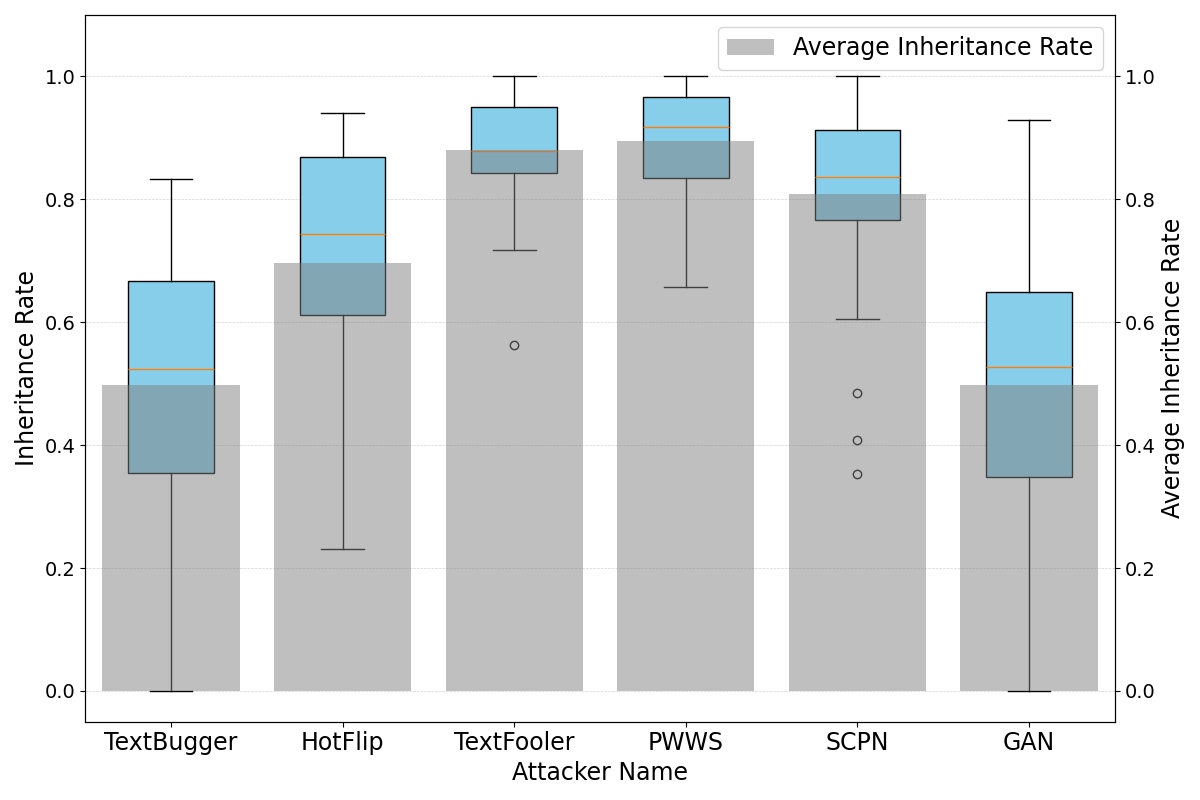}
\caption{The average transferable rates using different adversarial attack techniques (RQ3)}
\label{fig:robustness_transfer}
\end{figure}

Figure \ref{fig:robustness_transfer} shows the transferable rates using different adversarial attack techniques.
In general, the average transferable rate reaches 78.57\%, indicating that the vulnerable samples could be transferred along the model chains to some extent on HF.
In particular, the transferable rates exceed 60\% for 4/6 adversarial attack approaches, i.e., HotFlip, TextFooler, PWWS, and SCPN.
This indicates that, for most techniques, if original samples could successfully attack the upstream model, they are more likely to be vulnerable to the corresponding downstream model.

This transitivity suggests that fine-tuning has not enhanced the downstream models' resistance to vulnerabilities known in upstream models and may even perpetuate or amplify these vulnerabilities within the model chains. This is especially critical for popular models frequently used for further fine-tuning, whose inherent vulnerabilities could impact a wide range of downstream applications. 
Therefore, developers should pay more attention to these potential adversarial risks when fine-tuning models.

\begin{tcolorbox}[colback=gray!50!white]
\textbf{Finding 5}:
\textit{
The vulnerability embodying in vulnerable samples could transfer along the model chains on HF with an average rate of  78.57\%.
When users are preparing to fine-tune based on open-source models, it is necessary to pay attention to the inherent robustness risks of upstream models, as they are likely to be transmitted to downstream applications.}
\end{tcolorbox}

%% file: 7.implication.tex
\section{Implications}

In this section, we'd like to elaborate more on the implications of our empirical studies for readers of interest to take away.

\textbf{The guidance of selecting proper reuse manner for security-critical applications.}
Given the open-source models, users typically reuse them in two manners: (1) directly apply them in applications without modifications and (2) fine-tune them to suit current application scenarios better.
Based on the results in RQ2 and RQ3, the two manners suffer different adversarial risks.
Therefore, disregarding the considerations of data and computational resources required for fine-tuning, users need to comprehend the model's architecture to be reused and its regularity of robustness change during the fine-tuning process to make the proper choice in the security-critical applications.
For instance, for open-source models with BERT architecture, selecting the most appropriate models from vast and diverse models on HF for direct applications would be more secure, as fine-tuning would degrade the adversarial robustness.
On the contrary, fine-tuning is the best way to reuse the models with Electra architecture to resist adversarial attacks.

\textbf{The significance of vulnerable samples for targeted defenses.}
In our empirical study, the generated adversarial samples can not only be used to assess the adversarial robustness of open-source models but also hold significance for targeted defense approaches (e.g., adversarial sample detection and adversarial training) based on therein vulnerable samples.
For example, adversarial sample detection techniques often involve training a binary classification model to predict whether an input is an adversarial sample. 
The adversarial samples generated in our study, especially the vulnerable samples that successfully attack the models, can be used as training data to improve the performance of detection models, achieving targeted defenses. 
Additionally, vulnerable samples can be used for data augmentation, enhancing the target model through adversarial training.

\textbf{Leverage the upstream model’s vulnerable samples for a more targeted attack on the downstream models.}
From the finding drawn in Section \ref{subsubsec_RQ3.2}, it is evident that if we have access to the vulnerable samples of the upstream model, namely the data instances that successfully attack the upstream model, then utilizing these samples to attack the downstream model can achieve an average success rate of 78.57\%, which is 20.33\% (78.57\% vs. 58.24\%) higher compared to using a standard dataset. 
This implies that the attacks on the upstream model can provide targeted information, leading to improved success rates in attacking the downstream model.

\textbf{Inspiration from Electra architecture for robustness enhancement.}
In Section \ref{subsubsec_RQ3.1}, we find that models with Electra architecture demonstrate robustness improvement after fine-tuning. 
Electra employs a replaced token detection (RTD) strategy, involving a generator and a discriminator; the generator is responsible for replacing tokens in the sentence, while the discriminator determines whether each token is a replacement.
The RTD strategy employed in Electra architecture may lead to robustness improvement due to its similarity to adversarial attack operations (namely replacing tokens to create disturbances). 
Electra, by training the model to recognize these minor changes, inadvertently enhances the model's resistance to such disruptions.
Inspired by this observation, targeted considerations for adopting techniques aligned with replacing tokens can also be explored to implement robustness enhancement.

%% file: 8.threat.tex
\section{Threats to Validity}

\textbf{External Validity.}
The external threats are related to the generalization of the empirical studies.
First, our subject models are collected from one AI model hub, i.e., HF. 
The relevant findings may differ for platforms like TensorFlow Hub and PyTorch Hub.
However, HF is one of the representative platforms attracting substantial attention and usage.
The empirical study based on HF is still of critical guiding significance.
Second, our study focuses on text classification models, and the results may not be generalized to other tasks, such as text generation and image classification.
Even so, text classification is a vital AI task widely used in many scenarios, such as intelligence analysis, social content analysis, and news classification. 
Related research also holds broad application prospects.
Third, the robustness of models and their fine-tuning are assessed through six adversarial attack techniques and four datasets.
The results may differ under other attack techniques and datasets.
However, the techniques and datasets employed are all prevalent and representative.
Additionally, they are used for robustness assessment in previous studies \cite{raman2023modeltuning,morris2020textattack,si2021benchmarking}, which could alleviate the threat.

\textbf{Internal Validity.}
The internal threats relate to experimental errors and biases.
First, we employ the ChatGPT to identify the upstream model names from the descriptions in ``Model Card''.
The bias of ChatGPT may be introduced in our study.
However, ChatGPT is a widely recognized commercial system that performs well in information extraction tasks~\cite {Bo2023entity}.
Furthermore, we evaluate the identification performance with a sampled dataset (details in Section \ref{sec:chain_cunstruction}), and 97\% accuracy reflects its credibility.
Second, for two adversarial attack techniques, TextBugger and GAN, there is randomness when generating adversarial samples according to their technical principles.
Therefore, the attack performance is unstable, which may threaten the experimental results.
To alleviate the threat, we run the attack technique three times for each experimental setting, and the average performance is utilized. 

%% file: 9.rw.tex
\section{Related Work}
\subsection{AI Model Reuse}
For AI models, the development is a resource-intensive process involving complex algorithm design, long-time computational resource consumption, and potential data labeling.
Reusing open-source models provides a promising way to build effective task-specific models with relatively lower costs.
Correspondingly, techniques and empirical studies for reusing AI models in their reusability, challenges, and respective improvement methods have emerged.
Qi et al. proposed a tool called SEAM, which is specifically designed to redesign trained deep neural network models to improve their reusability~\cite{qi2023reusing}.
Pan et al. conducted an empirical study to analyze the errors that occur when reusing pre-trained NLP models, and proposed strategies to reduce these errors~\cite{pan2022empirical}. 
Jiang et al. systematically evaluated the decision-making process for reuse of pre-trained models, and clarified the key attributes and challenges in reuse~\cite{jiang2023empirical}. 
Taraghi et al. conducted a mixed-method empirical study on pre-trained model reuse on HF, analyzed the challenges faced by users and the trend of model reuse, and proposed relevant guidance suggestions~\cite{taraghi2024deep}.
Zhao et al. developed a query framework called MMQ, which optimizes the model reuse process, enabling users to more accurately select the pretrained models needed for their tasks~\cite{zhao2023finding}.
Davis et al. explored the challenges and future direction of deep neural network reuse, and discussed in detail different types of reuse methods~\cite{davis2024reusing}. 
Together, these studies advance the theory and practice of AI model reuse. 
However, despite the above empirical studies on model reuse from various perspectives, there are fewer studies on the adversarial robustness of the open-source AI models, especially their fine-tuning chains.
Our empirical study can fill this gap.

\subsection{Empirical Studies On Hugging Face}
Since its inception, HF has become a pivotal platform in the NLP community. 
This platform not only offers a rich repository of open-source AI models but also fosters the sharing of open science and technology. 
Meanwhile, HF has attracted substantial empirical studies in the literature. 
Mitchell et al. introduced the concept of Model Cards, which has been used as an essential attribute of maintained models on HF, a standardized document designed to increase the transparency of AI models by detailing their performance and evaluation processes to promote the responsible use of models and a better understanding of their behavior~\cite{mitchell2019model}. 
Ait et al. evaluated the feasibility of using the Huggingface Hub as a data source for empirical research by analyzing its features~\cite{ait2023suitability}.
McMillan-Major et al. demonstrate how the HF platform and the GEM (Geospace Environment Modeling) workshop can be used to implement and customize standardized documentation templates for Model Cards and Data Cards to improve the quality and transparency of documentation of datasets and models in NLP and generative domains~\cite{mcmillan2021reusable}.
Castaño et al. explore the dynamics of community engagement, model evolution, and maintenance, providing crucial insights into future model development strategies~\cite{castaño2024analyzing}. 
Jiang et al. analyzed the naming conventions of pretrained models on the Hugging Face platform and developed an automated tool named DARA to detect anomalies in naming \cite{jiang2024naming}.
Yang et al. analyzed the current state of AI dataset documentation on the Hugging Face platform, the community's usage of dataset cards, and proposed guidelines for writing effective dataset cards~\cite{yang2024navigating}.
Taraghi et al. examine the challenges and trends in reusing pre-trained models in the community, highlighting the problems users face in understanding and applying models and proposing strategies to address them~\cite{taraghi2024deep}. 
Castaño et al. focus on carbon emissions during model training on the platform and promote initiatives for green model development~\cite{Casta_o_2023}. 

%% file: 10.conclusion.tex
\section{Conclusion}
To reuse the open-source models on HF, it is critical to understand their adversarial robustness against potential human interference to ensure the security of domain models in practical applications. 
This paper aims to investigate the adversarial robustness of open-source AI models and their fine-tuning chains, to provide insights into the potential adversarial risks.
Empirical analysis on the text classification models shows that, under the prevalent circumstance of model reuse, open-source text classification models exhibit certain deficiencies in adversarial robustness.
Furthermore, inherent robustness vulnerability in the text classification models exhibits a certain degree of transitivity in the fine-tuning chains, which highlights the security concerns for security-sensitive users in the downstream applications.
In the future, we are going to explore the adversarial robustness of more types of open-source AI models and design automated methods for targeted defense or robustness enhancement.